\documentclass[aps,prd,twocolumn,showpacs,superscriptaddress,groupedaddress,amsmath]{revtex4}  
\usepackage{graphicx}  
\usepackage{dcolumn}   
\usepackage{bm}        
\usepackage{amssymb}   
\usepackage{natbib}
\usepackage{color}
\usepackage{ulem}
\usepackage{epsfig}
\usepackage{float}
\usepackage{mathrsfs}
\usepackage{amsmath}

\newcommand{\mnras}{Mon.~Not.~R.~Astron.~Soc.}

\newcommand{\phrd}{Phys.~Rev.~D.}
\newcommand{\be}{\begin{equation}}
\newcommand{\ee}{\end{equation}}
\newcommand{\bea}{\begin{eqnarray}}
\newcommand{\eea}{\end{eqnarray}}

\begin{document}

\widetext


\title{Averaged universe confronted with cosmological observations: A fully covariant approach}
\author{Tharake Wijenayake}
\author{Weikang Lin}
\author{Mustapha Ishak\footnote{mishak@utdallas.edu}}
\affiliation{Department of Physics, The University of Texas at Dallas, Richardson, TX 75083, USA}
\date{\today}

\begin{abstract}
One of the outstanding problems in general relativistic cosmology is that of the averaging. That is, how the lumpy universe that we observe at small scales averages out to a smooth Friedmann-Lemaitre-Robertson-Walker (FLRW) model. The root of the problem is that averaging does not commute with the Einstein equations that govern the dynamics of the model. This leads to the well-know question of backreaction in cosmology. In this work, we approach the problem using the covariant framework of Macroscopic Gravity (MG). We use its cosmological solution with a flat FLRW macroscopic background where the result of averaging cosmic inhomogeneities has been encapsulated into a backreaction density parameter denoted $\Omega_\mathcal{A}$. We constrain this averaged universe using available cosmological data sets of expansion and growth including, for the first time, a full CMB analysis from Planck temperature anisotropy and polarization data, the supernovae data from Union 2.1,  the galaxy power spectrum from WiggleZ, the weak lensing tomography shear-shear cross correlations from the CFHTLenS survey and the baryonic acoustic oscillation data from 6Df, SDSS DR7 and BOSS DR9. We find that $-0.0155 \le \Omega_\mathcal{A} \le 0$ (at the 68\% CL) thus providing a tight upper-bound on the backreaction term. We also find that the term is strongly correlated with cosmological parameters such $\Omega_\Lambda$, $\sigma_8$ and $H_0$. While small, a backreaction density parameter of a few percent should be kept in consideration along with other systematics for precision cosmology.
\end{abstract}
\maketitle
\section{Introduction.}
The ongoing and planned high precision surveys aim at constraining cosmological models at the percent-level precision.
This is faced with a number of challenges from systematic effects in the data, computational frameworks, and theoretical issues.
It has become increasingly important to understand and take into account the nonlinear and relativistic effects of gravity. One such effect, that has garnered considerable attention in the literature, is the averaging problem in relativity and cosmology \cite{Clarksonetal}.
The problem stems from the fact that applying the Einstein's field equations to a smooth model of the universe and performing averaging to the observed lumpy universe are two operations that do no commute.
This non-commutation leads to a difference between the dynamics of the exact homogeneous and isotropic Friedmann-Lemaitre-Robertson-Walker (FLRW) model and the effective model obtained from an averaged universe. This difference is referred to as the backreaction
and enters into the evolution equations of the models. While it is fair to assert that one would not expect huge effects from such a backreaction, it remains an open question whether its effect can be at the same level as that of other systematics in the data and thus worth considering in the quest of precision cosmology.
Various schemes of averaging have been proposed in the literature and can be found in, for example, the review \cite{Clarksonetal}. The authors of Ref. \cite{Buchert:2003} developed a formalism based on spatial averaging of the scalar model evolution equations which attracted a lot of attention. For example, recently the formalism has been generalized in LRS class II spacetimes \cite{2015Kavspar-Svitek}. A fully covariant approach that can be used to average tensors was proposed in \cite{Zalaletdinov} and is considered in the field among the most promising formalisms to study the averaging problem. In this approach, the averaging produces a macroscopic cosmological model so the formalism is referred to as macroscopic gravity (MG). We use this formalism in this work.

In this paper, we employ an elaborate framework that includes expansion and growth observable functions developed for the MG averaged universe. Some of the details of the framework are described in our previous extended paper \cite{2015PhRvD..91f3534W}. Here we apply it to perform for the first time a full cosmic  microwave background analysis using temperature anisotropies and polarization data from Planck \cite{2015arXiv150201589P}. We also use the galaxy power spectrum data from WiggleZ \cite{2010MNRAS.401.1429D, 2012PhRvD..86j3518P}, the baryonic acoustic oscillations (BAO) from 6Df \cite{2011MNRAS.416.3017B}, SDSS DR7 \cite{2015MNRAS.449..835R} and BOSS DR11 \cite{2014MNRAS.441...24A}, distances to supernovae from the Union 2.1 compilation  \cite{Suzuki:2011hu} and the weak lensing tomography shear-shear cross correlations from the CFHTLenS survey \cite{2012MNRAS.427..146H}. We use  modified versions of the publicly available Boltzmann code CAMB \cite{2000ApJ...538..473L} and the Markov chain Monte Carlo code CosmoMC \cite{Lewis:2002ah}.

\section{Averaged universe and macroscopic gravity formalism.}
MG is a fully covariant, non-perturbative,  effective model which describes  the large scale behavior of space-time  \cite{Zalaletdinov}. The formalism uses a bivector $\mathcal{A}^{\alpha}_{\,\,\,\,\alpha'}(x,x')$ to define covariant averages of tensor fields. The average of an arbitrary tensor field $P^{\alpha ...}_{\,\,\,\,\,\,\,\,\beta ...}$ at a point $x$ over some averaging region  $ \Sigma_{x}$ is defined as
\be
\resizebox{0.5\textwidth}{!}{$\overline{P^{\alpha ...}_{\,\,\,\,\,\,\,\,\beta ...}}(x)  =\frac{1}{V_{\Sigma_{x}}} \int_{\Sigma_{x}} P^{\alpha' ...}_{\,\,\,\,\,\,\,\,\beta' ...}(x') \mathcal{A}^{\alpha}_{\,\,\,\,\alpha'}(x,x')\mathcal{A}^{\beta'}_{\,\,\,\, \beta}(x',x) ...  \sqrt{-g(x')}d^4x'        $}             \label{avefield}
\ee
where, $ V_{\Sigma_{x}} = \int_{\Sigma_{x}} \sqrt{-g(x')}d^4x'$ is the 4-volume of the averaging region for support point x, and the averaging bivectors satisfy the conditions  $\lim_{x'  \to x} \mathcal{A}^{\alpha}_{\,\,\,\,\beta'}(x',x) =\delta^{\alpha}_{\beta}$, $\mathcal{A}^{\alpha}_{\,\,\,\,\beta'}(x,x')\mathcal{A}^{\beta'}_{\,\,\,\,\gamma''}(x',x'')=\mathcal{A}^{\alpha}_{\,\,\,\,\gamma''}(x,x'')$, $ \mathcal{A}^{\alpha'}_{\,\,\,\,[\beta, \gamma]} + \mathcal{A}^{\alpha'}_{\,\,\,\,[\beta, \delta'}\mathcal{A}^{\delta'}_{\,\,\,\, \gamma]}= 0 $ and $\mathcal{A}^{\alpha'}_{\,\,\,\, \beta ; \alpha'}   = 0$.

In the MG formalism the macroscopic affine connection is given by $\langle  {\mathscr{F}}^{\alpha}_{\,\,\,\, \beta \gamma}  \rangle$. Where ${\mathscr{F}}^{\alpha}_{\,\,\,\, \beta \gamma} $ is the ``bilocal extension" of the connection coefficients which transforms like a connection at $x$, like a scalar at $x'$, and reduces to the microscopic connection $\Gamma^{\alpha}_{\,\,\,\, \beta \gamma}$ in the limit where $x'$ goes to $x$.
\be
{\mathscr{F}}^{\alpha}_{\,\,\,\, \beta \gamma} :=  \mathcal{A}^{\alpha}_{\,\,\,\, \epsilon'} \left(        \mathcal{A}^{ \epsilon'}_{\,\,\,\, \beta ,\gamma} + \mathcal{A}^{ \epsilon'}_{\,\,\,\, \beta ;\sigma'}\mathcal{A}^{\sigma'}_{\gamma}   \right)
\ee
Angle brackets denote integration over the averaging region divided by the volume of the averaging region, for example
 $\overline{P^{\alpha}} \equiv \langle \mathcal{A}^{\alpha}_{\;\;\alpha'}P^{\alpha'} \rangle$.
There is a macroscopic curvature tensor ($M^{\alpha}_{\,\,\,\, \beta \gamma \delta}$) and a macroscopic metric ($G_{\alpha \beta}$) corresponding to the macroscopic connection ($\langle{{\mathscr{F}}^{\alpha}_{\,\,\,\, \beta \gamma} }\rangle$). Additionally there is a connection ($\Pi^{\alpha}_{\,\,\,\, \beta \gamma}$) corresponding to the averaged microscopic Riemann tensor (${\bar{R}}^{\alpha}_{\,\,\,\, \beta \gamma \delta}$).
The difference between the two connection coefficients is defined as the  affine deformation tensor  (${A}^{\alpha}_{\,\,\,\, \beta \gamma}= \,\langle{{\mathscr{F}}^{\alpha}_{\,\,\,\, \beta \gamma} }\rangle-\Pi^{\alpha}_{\,\,\,\, \beta \gamma}$).

Under the assumption of the splitting rules $\langle{\mathscr{F}}^{\alpha}_{\,\,\,\, \beta \gamma} g_{\mu \nu} \rangle\,  =  \langle{\mathscr{F}}^{\alpha}_{\,\,\,\, \beta \gamma}\rangle\bar{g}_{\mu \nu} $ and  $\langle{\mathscr{F}}^{\alpha}_{\,\,\,\, \beta \gamma} {\mathscr{F}}^{\delta}_{\,\,\,\, \epsilon \eta}g_{\mu \nu} \rangle \, =  \langle{\mathscr{F}}^{\alpha}_{\,\,\,\, \beta \gamma}\rangle \langle{\mathscr{F}}^{\delta}_{\,\,\,\, \epsilon \eta}\rangle \bar{g}_{\mu \nu} $,   the Einstein field equation (EFE), and Cartan's structure equations  can be averaged out to construct the MG field equations
\be
\resizebox{0.5\textwidth}{!}{$\bar{g}^{\beta \epsilon} M_{\beta \gamma} -\frac{1}{2}\delta^{\epsilon}_{\gamma}\bar{g}^{\mu \nu}M_{\mu \nu}  =8\pi G \left[  \bar{T}^{\epsilon}_{\gamma}   -\left(Z^{\epsilon}_{\,\,\,\, \mu \nu \gamma} + \frac{1}{2}\delta^{\epsilon}_{\gamma}Q_{\mu \nu}         \right)  \bar{g}^{\mu \nu}  \right] $}  \label{aveEFE}
 \ee
 \be
 \bar{R}^\alpha_{\,\,\,\,\beta [ \sigma \rho ; \lambda]}  = A^\epsilon_{\,\,\,\, \beta  [ \rho} \bar{R}^\alpha_{\,\,\,\,  \underline{\epsilon} \sigma \lambda] }+A^\alpha_{\,\,\,\, \epsilon  [ \rho} \bar{R}^\epsilon_{\,\,\,\,  \underline{\beta} \sigma \lambda] }
\ee

\be
M^\alpha_{\,\,\,\, \beta \rho \sigma}=\bar{R}^\alpha_{\,\,\,\, \beta \rho \sigma}+Q^\alpha_{\,\,\,\, \beta \rho \sigma}
\ee

The model is completely specified by four tensor potentials which depend on the inhomogeneous substructure and the averaging scale. These are the affine deformation tensor ${A}^{\alpha}_{\,\,\,\, \beta \gamma}$ defined above, the correlation 2-form $Z^{\alpha \,\,\,\, \,\,\,\, \mu}_{ \,\,\,\,\beta[\gamma \,\,\,\,\underline{\nu}\sigma]}$, the correlation 3-form $Y^{\alpha \,\,\,\, \,\,\,\, \mu \,\,\,\, \,\,\,\, \theta}_{\,\,\,\, \beta [ \gamma \,\,\,\, \underline{\nu} \sigma \,\,\,\, \underline{\kappa} \pi]}$ and the correlation 4-form  $X^{\alpha \,\,\,\, \,\,\,\, \mu \,\,\,\, \,\,\,\, \theta \,\,\,\,  \,\,\,\,  \tau}_{\,\,\,\, \beta [ \gamma \,\,\,\, \underline{\nu} \sigma \,\,\,\, \underline{\kappa} \pi \,\,\,\,  \phi \psi]}$.
The correlation 2-form is defined as :
\be
  Z^{\alpha \,\,\,\, \,\,\,\, \mu}_{ \,\,\,\,\beta[\gamma \,\,\,\,\underline{\nu}\sigma]} \;:=\; \langle{ \mathscr{F}}^{\alpha }_{ \,\,\,\,\beta[\gamma}{ \mathscr{F}}^{\mu }_{ \,\,\,\,\underline{\nu}\sigma]}\rangle  \;-\; \langle{ \mathscr{F}}^{\alpha }_{ \,\,\,\,\beta[\gamma}\rangle \langle{ \mathscr{F}}^{\mu }_{ \,\,\,\,\underline{\nu}\sigma]}\rangle  \nonumber
\ee
and has the following contractions:
\begin{align}
 Q^{\alpha}_{\,\,\,\,\beta \rho \mu} &= -2 Z^{\epsilon \,\,\,\, \,\,\,\, \alpha}_{ \,\,\,\,\beta\rho \,\,\,\,\epsilon\gamma}     &      Z^{\epsilon}_{\,\,\,\, \mu \nu \gamma}&=2Z^{\epsilon \,\,\,\, \,\,\,\, \delta}_{ \,\,\,\,\mu\delta \,\,\,\,\nu\gamma}  \nonumber \\
 & Q_{\mu \nu}=Q^{\epsilon}_{\,\,\,\, \mu \nu \epsilon}=Z^{\delta}_{\,\,\,\, \mu \nu \delta}
 \end{align}
The differential properties of the correlation 2-form are set by the correlation 3-form and the correlation 4-form. For the simplest model, the correlation 3 and 4-forms are zero and the following differential and algebraic equations hold.
The differential cyclic constraint
\be
Z^{\alpha \,\,\,\, \,\,\,\, \mu}_{ \,\,\,\,\beta[\gamma \,\,\,\,\underline{\nu}\sigma ||\lambda]}=0 \label{dz}
\ee
The integrability condition
 \be
 \begin{split}
Z^{\epsilon \,\,\,\, \,\,\,\, \gamma}_{ \,\,\,\,\beta[\mu \,\,\,\,\underline{\delta}\nu}M^{\alpha}_{\,\,\,\, \underline{\epsilon} \kappa \pi]}   - Z^{\alpha \,\,\,\, \,\,\,\, \gamma}_{ \,\,\,\,\epsilon[\mu \,\,\,\,\underline{\delta}\nu}M^{\epsilon}_{\,\,\,\, \underline{\beta} \kappa \pi]}   \\+Z^{\alpha \,\,\,\, \,\,\,\, \epsilon}_{ \,\,\,\,\beta[\mu \,\,\,\,\underline{\delta}\nu}M^{\gamma}_{\,\,\,\, \underline{\epsilon} \kappa \pi]}   - Z^{\alpha \,\,\,\, \,\,\,\, \gamma}_{ \,\,\,\,\beta[\mu \,\,\,\,\underline{\epsilon}\nu}M^{\epsilon}_{\,\,\,\, \underline{\delta} \kappa \pi]}   = 0  \label{zm}
  \end{split}
\ee
The quadratic constraint equation
\be
\begin{split}
Z^{\delta \,\,\,\, \,\,\,\, \theta}_{ \,\,\,\,\beta[\gamma \,\,\,\,\underline{\kappa}\pi}Z^{\alpha \,\,\,\, \,\,\,\, \mu}_{ \,\,\,\,\underline{\delta}\epsilon \,\,\,\,\underline{\nu}\sigma]}   +   Z^{\delta \,\,\,\, \,\,\,\, \mu}_{ \,\,\,\,\beta[\gamma \,\,\,\,\underline{\nu}\sigma}Z^{\theta \,\,\,\, \,\,\,\, \alpha}_{ \,\,\,\,\underline{\kappa}\pi \,\,\,\,\underline{\delta}\epsilon]}  \\+    Z^{\alpha \,\,\,\, \,\,\,\, \delta}_{ \,\,\,\,\beta[\gamma \,\,\,\,\underline{\nu}\sigma}Z^{\mu \,\,\,\, \,\,\,\, \theta}_{ \,\,\,\,\underline{\delta}\epsilon \,\,\,\,\underline{\kappa}\pi]}
+  Z^{\alpha \,\,\,\, \,\,\,\, \mu}_{ \,\,\,\,\beta[\gamma \,\,\,\,\underline{\delta}\epsilon}Z^{\theta \,\,\,\, \,\,\,\, \delta}_{ \,\,\,\,\underline{\kappa}\pi \,\,\,\,\underline{\nu}\sigma]} \\ +  Z^{\alpha \,\,\,\, \,\,\,\, \theta}_{ \,\,\,\,\beta[\gamma \,\,\,\,\underline{\delta}\epsilon}Z^{\mu \,\,\,\, \,\,\,\, \delta}_{ \,\,\,\,\underline{\nu}\sigma \,\,\,\,\underline{\kappa}\pi]}
  +Z^{\alpha \,\,\,\, \,\,\,\, \delta}_{ \,\,\,\,\beta[\gamma \,\,\,\,\underline{\kappa}\pi}Z^{\theta \,\,\,\, \,\,\,\, \mu}_{ \,\,\,\,\underline{\delta}\epsilon \,\,\,\,\underline{\nu}\sigma]}  =  0   \label{zz}
  \end{split}
\ee
\\
\section{Cosmological solutions to MG and the Friedmann macroscopic universe.}
The MG equations have been solved for the case when the macroscopic metric is given by a flat FLRW metric \cite{Coley:2005ei,vandenHoogen:2009nh,Clifton:2012fs} under the assumption that the correlation 3-form and 4-form are zero, the correlation 2-form  ($Z^{\alpha \,\,\,\, \,\,\,\, \mu}_{ \,\,\,\,\beta\gamma \,\,\,\,\nu\sigma}$) and the affine deformation tensor (${A}^{\alpha}_{\,\,\,\, \beta \gamma}$) are invariant under the six parameter group of Killing vectors (corresponding to the three translational and three rotational symmetries of the metric), and the electric part of the correlation tensor is zero ($Z^{\alpha \,\,\,\, \,\,\,\, \mu}_{ \,\,\,\,\beta\gamma \,\,\,\,\nu\sigma} u^{\sigma}    = 0$ where $u^{\sigma}$ is the timelike vector orthogonal to the hypersurface of homogeneity).

With an averaged stress energy tensor of the form of a perfect fluid, the macroscopic  EFE (\ref{aveEFE}) gives:
\be
\frac{\dot{a}^2}{a^2} = \frac{8\pi G}{3}\rho -\frac{1}{3}\frac{\mathcal{A}^2}{a^2} +\frac{\Lambda}{3}
\ee
\be
\frac{2\ddot{a}}{a}+\frac{\dot{a}^2}{a^2}  = -8 \pi Gp - \frac{1}{3}\frac{\mathcal{A}^2}{a^2} +\Lambda
\ee
\noindent where dots denote partial differentiation with respect to the time coordinate $t$, and $\mathcal{A}^2$ is a positive constant that implicitly depends on the underlying inhomogeneous structure and the averaging scale. The sign of this constant  derives from the symmetry condition on the affine deformation tensor under the group of killing vectors, given in the previous paragraph \cite{vandenHoogen:2009nh, 2015PhRvD..91f3534W}. This also sets the sign of $\Omega_\mathcal{A}$ defined further below, it can be viewed as a mathematical and physical prior. Hence, the macroscopic gravity correlations appear like an extra positive spatial curvature term in the Friedmann's equations. Integrating the null geodesic equation in a flat FLRW macroscopic metric we get the luminosity distance relation modified only by the change in expansion history due to the backreaction term,
\begin{equation}
d_{L}  = \frac{1}{a H_0} \int^{a}_{a'=1} \frac{da'}{{(\Omega_{\mathcal{A}}{a'}^{2}     +\Omega_{\Lambda} {a'}^4+\Omega_{m} {a'} )}^{\frac{1}{2}}}.
\label{flat_mg_dl}
\end{equation}

\noindent where $\Omega_m\equiv\frac{8}{3} \pi G \rho_0 /H_{0}^2$ is the matter density parameter, $\Omega_{\Lambda}\equiv\Lambda /3H_{0}^2$ is the cosmological constant density parameter,  $\Omega_{\mathcal{A}}= -\mathcal{A}^2/{3 H_{0}^2 }$ is the backreaction density term also called ``gravitational energy" parameter due to averaging, and $H_{0}$ is the Hubble constant, all evaluated today.

\section{Growth of structures in a macroscopic gravity averaged universe.}
In order to describe the growth of structure in the macroscopic gravity model we can linearly perturb the macroscopic metric, the stress energy tensor and the MG potentials and solve the linearized MG field equations in order to obtain equations satisfied by the linear perturbations \cite{Clifton:2012fs, 2015PhRvD..91f3534W}.  In this situation, these long wavelength perturbations will describe the growth of structure at scales below the homogeneity scale but above the non-linear scales where the effects of the perturbations themselves will have significant impact on their dynamics. We note that the new evolution equations for these large-scale perturbations will account for the influence of the non-linear small-scale inhomogeneities that have been averaged over within the macroscopic gravity formalism.

The perturbed metric in the conformal Newtonian gauge takes the form
\begin{equation}
dS^2 =a(\eta)^{2} (-(1+2\phi)d\eta^{2}  +(1-2\psi)(dx^2 +dy^2 +dz^2))
\end{equation}

Assuming that the correlation 3-form and 4-form are zero at first order in the perturbations the MG field equations give:
\begin{eqnarray}
\nabla^2  \psi -3\mathcal{H}(\mathcal{H}\psi +\phi')  = 4 \pi G a^2( \delta \rho+ {\delta \rho}_{\mathcal{A}})\,\,\,\, \label{efe1} \\
\partial_i {(\mathcal{H}\phi +\psi')} =-4 \pi G a^2\left(p+\rho-\frac{2\mathcal{A}^2}{3a^2}\frac{1}{8\pi G}\right)\partial_i{\delta u}\,\,\,\,   \label{efe2} \\
\phi'' +\mathcal{H}\phi' +2\mathcal{H}\psi'+(2{\mathcal{H}}'+{\mathcal{H}}^2 )\phi + \nonumber \\
\frac{1}{3} \nabla^2(\psi-\phi) =4\pi G a^{2} \left(\delta p -\frac{{\delta \rho}_{\mathcal{A}}}{3}\right) \label{efe3} \\
\nabla^2(\psi-\phi) =8\pi G \Sigma\,\,\,\,    \label{efe4}
\end{eqnarray}
where $\delta\rho$ is the energy density perturbation,  $\delta p$ is the pressure perturbation, $\Sigma$ is the anisotropic stress$,\partial_{i}$ is the partial derivative, ${\delta \rho}_{\mathcal{A}}$ is the energy perturbation to the  gravitational stress energy tensor, a prime denotes the derivative with respect to $\eta$, $\partial_i \delta{u}$ is the irrotational part of the comoving peculiar velocity of the fluid and $\mathcal{H}$ is defined as $\frac{a'}{a}$.

The gravitational stress energy tensor takes the form of a perfect fluid with an effective equation of state $-1/3$.
\be
\rho_{\mathcal{A}}=-3\frac{\mathcal{A}^2}{a^2} + \delta \rho_{\mathcal{A}}+O(\epsilon ^2)    \label{grav1}
\ee
\be
p_{\mathcal{A}} =\frac{\mathcal{A}^2}{a^2}-\frac{1}{3} \delta \rho_{\mathcal{A}}+O(\epsilon ^2) \label{grav2}
\ee
where $\epsilon $ is the order of the linear perturbations.

 As noted in \cite{Clifton:2012fs} the MG equations also place constraints on the derivatives of  $\phi$ and $\delta{u_i} $ unless $\mathcal{A}^2 \sim  \epsilon$.
This seems to indicate that setting the correlation 3-form and 4-form to zero is not compatible with a perturbed macroscopic metric.  If we assume that the linear perturbations to the correlation 3-form and 4-form are of order $O(\epsilon)$, $\phi$ and $\delta{u_i} $ are no longer restricted but the gravitational stress energy tensor can have additional $O(\epsilon)$ corrections. In this work we did not consider such additional corrections assuming that the terms we used lead the corrections and hence the perturbed gravitational stress energy tensor takes the form of a perfect fluid with equation of state $-1/3$. While unlikely, it is unclear if these additional terms will affect significantly the observational constraints obtained on backreaction. Further studies should be devoted to deriving and analysing these additional terms. In this case the macroscopic EFE will be of the form given by Eq. (\ref{efe1}-\ref{efe4}), however the averaged stress energy tensor and the gravitational stress energy tensor are no longer conserved independently at first order.
The era of interest to us in this work is after matter-radiation equality. We will argue that deep within the matter dominated era (still well before decoupling) the perturbation to the gravitational energy density must be tightly coupled to the perturbation to the matter energy density, since it is the inhomogeneities and motion of the matter that cause the gravitational stress energy. In-fact we will consider the perturbation to the gravitational stress energy as being stationary in the matter frame (i.e. they have the same peculiar velocity).
Furthermore the matter energy is conserved at first order (even though the first order stress energy tensor is coupled to the first order gravitational stress energy tensor) since the proper mass is conserved.
Now, using stress energy conservation we can write the evolution equations for adiabatic perturbations.
\begin{equation}
\delta_{\mathcal{A}}=\frac{2}{3}\delta_m;\,\, \delta_m' = -\nabla^2 \delta u +3{\psi}'   \label{pert1}
\end{equation}
\begin{equation}
\begin{split}
\left(         1-\frac{2}{3} \frac{\Omega_{\mathcal{A}}}{\Omega_m}\right) \left(   {\nabla^2 \delta u}' -\nabla^2\psi \right) +\left(         1-\frac{4}{3} \frac{\Omega_{\mathcal{A}}}{\Omega_m}\right)\mathcal{H}\nabla^2 \delta u \\ +\frac{1}{6}\frac{\Omega_{\mathcal{A}}}{\Omega_m}\nabla^2 \delta_m =0  \label{pert3}
\end{split}
\end{equation}

\begin{table*}
\caption{\label{tab:1} Marginalized parameter constraints (68\% confidence) from the cosmological observations, the CMB temperature  (Planck) and low l polarization data {(lowP)} from Planck 2015, the distance to supernovae data from Union 2.1 {(Sn)}, the  galaxy power spectrum from WiggleZ {(MPK)}, the weak lensing tomography shear-shear cross correlations from the CFHTLenS survey {(wl)} and the baryonic acoustic oscillation data from 6Df, SDSS DR7 and BOSS DR9 {(BAO)}.}
\begin{ruledtabular}
\begin{tabular}{ccccccc}

Parameters &\multicolumn{2}{c}{Planck}& \multicolumn{2}{c}{Planck+lowP+Sn+MPK+wl}& \multicolumn{2}{c}{Planck+lowP+Sn+MPK+wl+BAO}\\ [0.5ex]

& MG&Vanilla&MG	& Vanilla&MG	& Vanilla	\\
\hline
\\
$\Omega_b h^2$&	$0.02216 \pm 0.00023$&    	$ 0.02220\pm 0.00023$&                  $0.02223 \pm 0.00022$&   	 $ 0.02231\pm 0.00022$&    $ 0.02218\pm 0.00021$&    $0.02228 \pm 0.00020$\\
\\
$\Omega_c h^2$&	$0.1201 \pm 0.0022$&    		$ 0.1198\pm 0.0022$&                              $0.1188 \pm 0.0018$&   	 $ 0.1178\pm 0.0017$&    $0.1194 \pm 0.0015$&    $ 0.1182\pm 0.0011$\\
\\
$\theta$&			$1.04080 \pm 0.00049$&    	$ 1.04086\pm 0.00048$&  				  $ 1.04093\pm0.00046 $&   	 $ 1.04109\pm 0.00044$&    $ 1.04086\pm 0.00042$&    $ 1.04104\pm 0.00041$\\
\\
$\tau$&		$0.079 \pm 0.019$&    				$ 0.079\pm 0.019$&   		 $0.066 \pm 0.015$&   	 $ 0.067\pm 0.015$&    $0.059 \pm 0.013$&    $ 0.065\pm 0.013 $\\
\\
$\Omega_\mathcal{A}$&    $    -0.0234_{-0.0051}^{+0.0234}$&    	$N/A$&    					$ -0.0196^{+0.0196}_{-0.0053} $&    	$ N/A$&    $ -0.0124_{-0.0031}^{+0.0124}$&    $N/A$\\
\\
$\log{A_s}$& 		$ 3.092  \pm 0.037$&                      	$ 3.091\pm 0.036$&    		$3.062 \pm 0.028$&    $3.06 \pm 0.028$&    $3.05 \pm 0.025$&    $ 3.06\pm0.024 $\\
\\
$n_s$& 	$0.9642 \pm 0.0064$&    			$0.9648\pm 0.0064$&  					$0.9670 \pm0.0057 $&    $ 0.969\pm 0.0054$&    $ 0.9653\pm 0.0049$&    $ 0.9683\pm 0.0043$\\
\\
$H_0$&	$69.3\pm 2.2$&   			 $ 67.3\pm 1.0$&   						 $ 69.6\pm 1.3$&    $68.13 \pm0.78 $&    $ 68.49\pm0.69 $&    $ 67.95\pm 0.52$\\
\\
$\Omega_{\Lambda}$&			$0.725\pm 0.038$&    $ 0.684\pm 0.014$&  				  $0.727 \pm 0.024$&    $0.696 \pm 0.010$&    $0.709 \pm 0.013$&    $0.6942 \pm 0.0068$\\
\\
$\Omega_m$& 	$0.298\pm 0.020$&    	$ 0.316\pm 0.014$&    				$ 0.293\pm 0.012$&    $ 0.304\pm0.010 $&    $0.3033 \pm0.0071 $&    $ 0.3058\pm 0.0068$\\
\\
$\sigma_8$&		$0.859\pm 0.031$&   		 $ 0.830 \pm 0.014$& 					   $0.838 \pm0.021 $&    $0.8133 \pm 0.0093$&    $0.826 \pm0.014 $&    $ 0.8126 \pm 0.0090$\\
\\
\end{tabular}
\end{ruledtabular}
\end{table*}

\begin{figure*}
\includegraphics[width=6.0in,height=6.0in,angle=0]{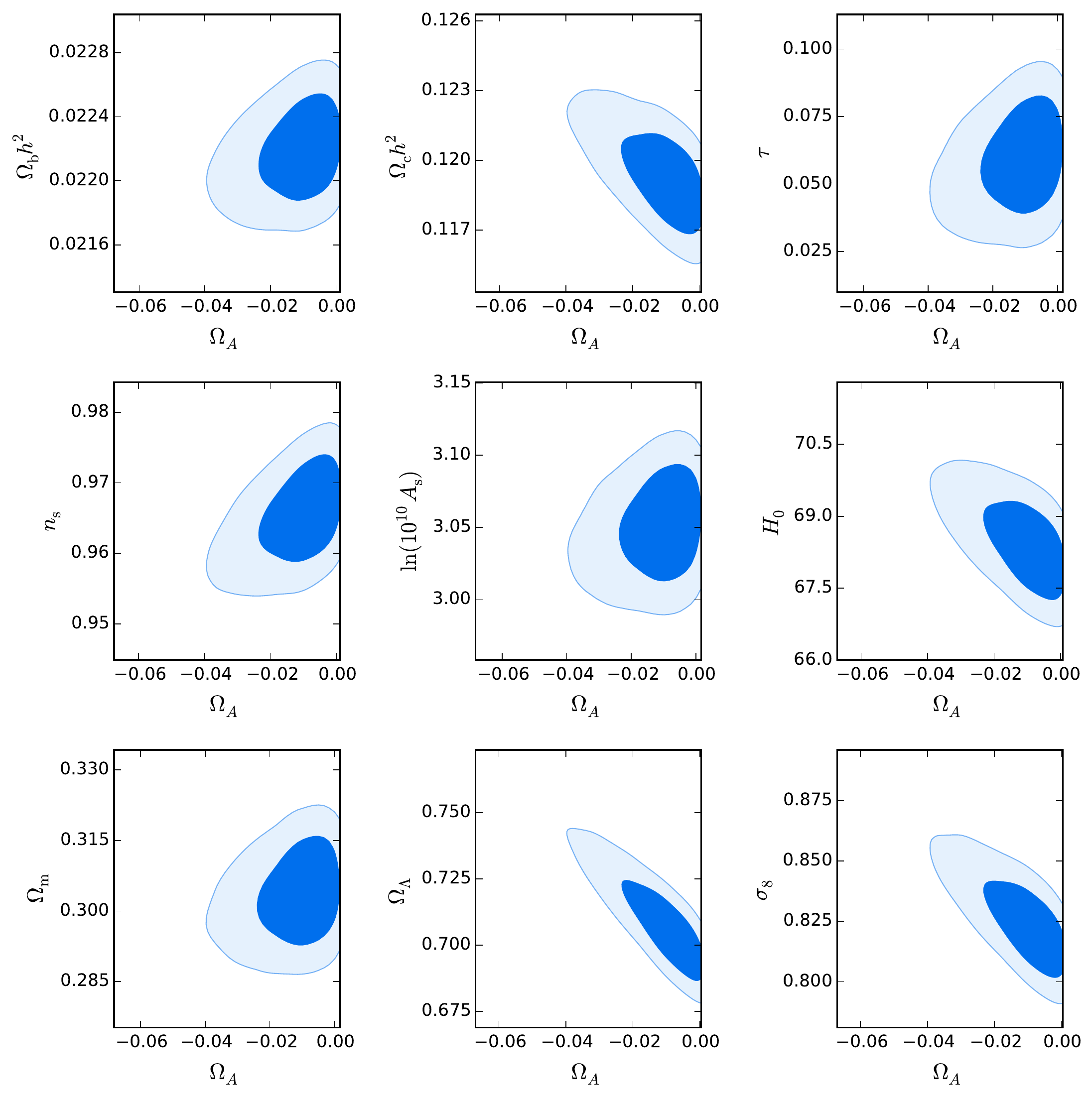}
\caption{\label{fig:1}  2-D  marginalized joint contour plots   (68\% and 95\% confidence levels) between $\Omega_{\mathcal{A}}$ and the other cosmological parameters for the Macroscopic gravity model from the combined data sets.}

\end{figure*}

\section{Resulting constraints and discussion.}
We fit the gravitational energy (backreaction) density parameter $\Omega_{\mathcal{A}}$ and the following six base cosmological parameters: the physical baryon density parameter $\Omega_{b}h^2$, the physical  cold dark matter parameter $\Omega_c h^2$, the ratio of the sound horizon to the angular diameter distance of the surface of last scattering $\theta$, the reionization optical depth $\tau$, the spectral index of the power spectrum $n_s$ and the logarithm of the amplitude of the primordial curvature power spectrum $\log{A_s}$. We use the MG framework and the data sets as described in the previous sections. We use the implementation and codes as described in our detailed previous paper \cite{2015PhRvD..91f3534W}.

Our results are summarized in Table \ref{tab:1} and Figure \ref{fig:1}. We find that the combined data sets constrain the averaging gravitational energy density parameter, or backreaction term to  $-0.0155 \le \Omega_\mathcal{A} \le 0$ (at the 68\% CL). This provides a tight constraint on backreaction within the MG formalism using the latest available data sets including a full CMB analysis.  We also find that including $\Omega_{\mathcal{A}}$ in the fit produces only small deviations of the six base cosmological parameters from their $\Lambda CDM$ values which are within one sigma. However, some derived parameters are highly degenerate with $\Omega_{\mathcal{A}}$. For example, $Corr(\Omega_{\mathcal{A}},\Omega_{\Lambda})=-0.86$, $Corr(\Omega_{\mathcal{A}},H_0)=-0.62$, $Corr(\Omega_{\mathcal{A}},\sigma_8)=-0.94$,  where $Corr(p,q)$ is the correlation coefficient between parameters $p$ and $q$.
These correlation coefficients are in agreement with the 2D confidence contours given in Fig. \ref{fig:1}. This significantly increases the errors associated with the constraints on these derived parameters.It is interesting that the well known tension between the local measurement of $H_0$ and the measurements from Planck  is not present in the MG model. Measurements from Cepheid and SN data by Reiss et al \cite{2011ApJ...730..119R} finds $H_0 =73.8\pm 2.4, $ and a reanalysis by  Efstathiou \cite{2014MNRAS.440.1138E} using a revised geometric maser distance to NGC4258 (one of the distance anchors)  finds $H_0=72.5 \pm2.5$.  The Planck value of $H_0 =67.3 \pm 1.0$ is off by two sigma while the MG Planck value is $H_0 =69.3 \pm 2.2$, within one sigma of both local measurements.
      	
It should be noted that the evolution equations for the perturbations (\ref{pert1}-\ref{pert3}) as used in our analysis hold only after the beginning of the matter dominated era, however we are interested here in backreaction effects after decoupling and well-after radiation domination so this should have no impact on the results. There are also some limitations in using some of the data sets. For example, when using the matter power spectrum, it is not possible to calculate the power spectrum for the non linear scales since there is no fitting model calibrated against simulations equivalent to the Halofit code for LCDM, however it is known that linear theory is a good fit to the data to $k \le 0.2h Mpc^-1$ \cite{2012PhRvD..86j3518P}. Using the Weak lensing data is more sensitive to this limitation but we verified that truncating the nonlinear scales has very little effect on the overall constraints obtained. Using and interpreting the BAO data depends on the growth of structure and hence the available data is for the $\Lambda CDM$ and would perhaps not hold if the effects of backreaction are significant. For the MG formalism, we assumed that the effect on light propagation is dominated by the contribution from the modified Friedmann's equation and we expect our main results here to hold even if other effects are introduced. {There are possible additional terms of the order $\epsilon$ from the perturbation that can be added to the gravitational stress energy tensor that we did not consider assuming that the terms we used capture the leading effect of backreaction. Although unlikely, it is unclear if such additional terms will change, in  a significant way,  the bounds obtained on backreaction from observations. The derivation and inclusion of such additional terms are beyond the scope of this paper and is left for further investigation. Moreover, we find that the constraints from the expansion history seems to provide the bulk of the constraining power on the backreaction term  \cite{2015PhRvD..91f3534W}.}
Finally, it is worth mentioning that the analysis here is based on the cosmological exact solution currently available in MG. It remains an open question if other viable cosmological solutions to MG exist and can be derived in the future.

There has been a considerable debate as to the significance of backreaction for cosmology, particularly regarding the framework of Green and Wald \cite{2011PhRvD..83h4020G,GreenWald2012,GreenWald2013,GreenWald2014} which is an extension of the Burnett's formulation of the shortwave approximation using weak limits to the problem of backreaction \cite{Burnett1989}. The main conclusion by Green and Wald is that if the matter stress energy satisfies the weak energy condition, the effective stress energy due to backreaction is trace-free and has the equation of state of radiation. They thus conclude that it has a negligible dynamical effect and cannot play the role of dark energy. In the MG formalism used in our paper, the effective stress energy tensor from averaging (backreaction) is not trace-free and has the equation of state of curvature.
Also, Green and Wald have found in \cite{2011PhRvD..83h4020G} that within their formalism the effect of small-scale inhomogeneities on the long wavelength perturbations is to add extra terms to the perturbed stress energy which they state corresponds to what one would expect from kinetic motions, and Newtonian potential energy and stresses.
Buchert et al. \cite{2015CQGra..32u5021B} objected that Green and Wald formalism is not general enough to capture generic properties of a realistic backreaction.
They also pointed out that some assumptions have discarded important points in the fitting and averaging problem, by construction. Green and Wald responded \cite{2015arXiv150606452G} to the criticism stating that their results remain valid and that there are various ways to define what backreaction is, which was in turn described again to be limited by  \cite{2015CQGra..32u5021B}. They finally proposed a heurestic description in \cite{GW2016heuristic} to justify their previous results.
Our work using MG formalism finds that current available data constrain the backreaction term to be small to have any large dynamical effect but still of possible significance at the level of systematic effects for precision cosmology.

We conclude from this cosmological analysis of the covariant MG formalism using multiple data sets and, for the first time, the full CMB analysis, that the effects of averaging inhomogeneities lead to a small backreaction term compared to other cosmological parameters.
Such a small backreaction term may remain significant when compared to percent-level systematics in the data for precision cosmology.

\acknowledgments
We thank S. M. Koksbang for comments on the manuscript, R. Zalaletdinov and A. Coley for comments on our previous related paper [5]. MI acknowledges that this material is based upon work supported in part by an award from the John Templeton Foundation and by the NSF under Grant 363 No. AST-1517768.

\end{document}